\documentclass[slac_one]{revtex4}
\usepackage{graphicx}
\usepackage{fancyhdr}
\begin{document}
\title{Charm spectroscopy beyond the constituent quark model} 

\author{J. Segovia, D.R. Entem, F. Fern\'{a}ndez}
\affiliation{Grupo de F\'{\i}sica Nuclear
Instituto de F\'{\i}sica Fundamental y Matem\'{a}ticas,
Universidad de Salamanca, 37008 Salamanca, Spain}

\begin{abstract}
We study the nature of the new charm and charmonium like states observed in the last few years, using a constituent quark model with an screened confinement potential. In the open charm sector we discuss different options for the $D_{sJ}(2860)$ including tetraquark configurations. In the hidden charm sector the new $Y(4360)$ and $Y(4660)$ are interpreted as $1^{--}$ states. The influence in the puzzling 3940 MeV region of multiquark configuration is discussed.   
\end{abstract}

\maketitle

\thispagestyle{fancy}

In the last few years the B factories (BaBar and Belle) and other machines have reported a large number of
new states with open and hidden charm: the puzzling $D_{sJ}^*$ states ($D_{sJ}^*(2317)$, $D_{sJ}(2460)$, $D_{sJ}(2860)$) and the XYZ ones
($X(3872)$, $X(3940)$, $Y(3940)$, $Z(3930)$, X(4160), $Y(4260)$, $Y(4360)$, $Y(4660)$ among others).
 
Some of them can be understood as $c\bar s$, or $c\bar c$ states while other shows unexpected properties being 
their energies and decay widths
quite different from those expected from quark potential
models. 
 
In this work we address this problem for the charm sector spectroscopy starting from a constituent quark model which has been able to describe the meson spectra from the light to the heavy sector ~\cite{vijb1}. 
The model is based on the assumption
that the light quark constituent mass appears as a consequence of the spontaneous
breaking of the original QCD $SU(3)_{L}\otimes SU(3)_{R}$ chiral symmetry at
some momentum scale, which is the most important nonperturbative phenomenon 
for hadron structure at low energies. In this domain 
of momenta, light quarks are quasiparticles with a constituent
mass interacting through scalar and pseudoscalar boson-exchange potentials. In the heavy quark sector chiral symmetry is explicitly bronken and this type of interaction does not act, however it constrains the model parameters through the ligth meson phenomenology.
Beyond the chiral symmetry breaking scale one expects the dynamics
being governed by QCD perturbative effects. They are taken into account
through the one-gluon-exchange potential. An important characteristic of the model is the use of a confining interaction taken
from lattice simulations as a linear potential screened at large distances due to quark-pair creation. This form of the potential is important to explain the huge degenerancy observed in the high excited light meson spectrum~\cite{seg} and turns out to be also important for the correct ordering of charmonium states.
Explicit expressions for 
the $qq$ and $q\bar q$ potentials are given in~\cite{vijb1}.

Physical mesons are easily identified with  $q\bar q$ states when loops are not important. This is the case of the pseudoscalar and vector mesons due to the P-wave nature of the loop dressing. On the contrary, in  P-wave state the contributing  quark loops are in an S-wave and  the coupling between $q\bar q$ pairs and $qq\bar q\bar q$ structures may be relevant. 
Thus, the inclusion of quarks loops (or tetraquarks) may influence $0^+$ and $1^+$ meson states like  
the $D_0^*(2308)$ and $D_{sJ}(2460)$. All negative 
parity and $2^+$ states should, in principle, be understood in terms only of $q\bar q$ components. 

The 1S and 1P excitations of the strange open-charm sector has been already discussed in Ref.~\cite{Vijb5}.
We enlarge this study focusing on the different options for the $D_{sJ}(2860)$ by studing its strong decays. 

Concerning the hidden charm sector
the screened confinement potential allow us to identify most of the $1^{--}$  states with definite $q\bar q$ pairs and its decay properties are established. We are able to indentify some of the states in the 3940 MeV puzzling region although for a complete description of all the states one needs to go beyond the $q\bar q$ model.

\section{THE $\bf{D_{sJ}(2860)}$ RESONANCE}
In 2006 the Babar collaboration reported a new state $D_{sJ}(2860)$ with a mass $2856.6\pm 5.0$ MeV and a 
width $\Gamma=48\pm 7\pm 10$ MeV. This state is observed in the $DK$ channel but no structure appears in the $D^*K$ invariant mass distribution in the same enrgy range. Thus we have  $J^P=0^+,1^-,2^+,3^-,...$ as possible quantum numbers. This state adds to the charm resonances reported during the last years like the $D_{sJ}(2317)$ and $D_{sJ}(2460)$ with masses typically 100 MeV lower than the quark model prediction. 
In an attempt to explain these later states, the authors of Ref.~\cite{Vijb5} included in the description of the open charm mesons tetraquark configurations. In constituent quark models  the more general wave 
function of a zero baryon number (B=0) hadron may be written as
$|\rm{B=0}>=\Omega_1\left|q\bar q\right>+\Omega_2\left|qq\bar q \bar q\right>+....$
To take into 
account simultaneously two-- and four--quark states the quark model hamiltonian  must be supplemented by a new term 
$H_{1}$, describing  the mixing between
$q\overline{q}$ and $qq\bar{q}\bar{q}$ configurations. In Ref.~\cite{Vijb5} 
 this hamiltonian is parametrized as $V_{q\overline{q} \leftrightarrow qq\bar{q}\bar{q}}=\gamma $ being $\gamma$ a constant.
 
When this parameter $\gamma$  is fixed to reproduce the 
mass of the $D_{sJ}^*(2317)$ meson 
they obtained the results shown in Table \ref{t1}. From these 
results one can appreciate that the description of the positive parity open-charm 
mesons improves, in particular the $D_{sJ}(2460)$,  when four--quark components are considered.

\begin{table}
\caption{Probabilities (P), in \%, of the wave function components 
and masses (QM), in MeV, of the open-charm mesons once the mixing between $q\bar q$ and $qq\bar q\bar q$ configurations 
is considered ~\cite{Vijb5}. Experimental data (Exp.) are taken from Ref. ~\cite{PDG}.} 
\label{t1}
\begin{tabular}{|c|cc||c|cc|}
\hline
\multicolumn{3}{|c||}{$I=0$~~~~~~$J^P=0^+$}    & \multicolumn{3}{|c|}{$I=0$~~~~~~$J^P=1^+$}  \\
\hline
QM                  	&2339   	&2847	&QM			&2421  		&2555 
\\
Exp.                	&2317.4$\pm$0.9	&$-$	&Exp.			&2459.3$\pm$1.3	&2535.3$\pm$0.6 \\
\hline
P($cn\bar s\bar n$) 	&28   		&55	&P($cn\bar s\bar n$)	&25  		&$\sim 1$  \\
P($c\bar s_{1^3P}$) &71   &25  &P($c\bar s_{1^1P}$)	&74  &$\sim 1$ \\
P($c\bar s_{2^3P}$) &$\sim 1$  &20  &P($c\bar s_{1^3P}$)&$\sim 1$ &98	\\
\hline
\end{tabular}
\end{table}

We have studied the spectrum and the decay properties of the open charm mesons in the 2800 MeV mass region. We found three states with masses 2873, 2883 and 2882 MeV (table~\ref{t2}) close to the experimental 
mass which correspond with the $1^-,2^-$ and $3^-$ $c\bar s$ $D$ $q\bar q$ waves. These has to be added to the 2847 MeV $0^+$ $c\bar s+cn\bar s\bar n$ 
(45\% and 55\% probability respectively) excitation obtained in Ref.~\cite{Vijb5}(table \ref{t2}).

To elucidate the nature of the $D_{sJ}(2860)$ we have calculated the strong decay widths of the two quarks states 
using the $^3P_0$ model. All of them have significant probability to decay through the $D^*K$ channel except the states $J^P=0^+$. As the $D^*K$ decay is not observed in the experimental data, the only remaining possibility for the $D_{sJ}(2860)$ state is the $0^+$ $c\bar s+cn\bar s\bar n$ state predicted in Ref~\cite{Vijb5}. Therefore one has to conclude that the puzzling situation of the $D_{sJ}$ meson states can be solved going beyond the $q\bar q$ constituent quark model.

\begin{table}
\caption{Predicted mass in MeV and strong width in MeV for the diferent $c\overline s$ states}
\label{t2}
\begin{center}
\begin{tabular}{|c|c|ccccc|}
\hline
$nL$ $J^P$& Mass & $DK$		&$D_s\eta $& $D^*K$  & $D^*_s\eta$   & $DK^*$\\
\hline								
$2P$ $0^+$&2966&   49.7 & 0.52    & 0.0     & 0.0           & 0.0    \\
$1D$ $1^-$&2873&50.1	  & 1.2     &36.9     &	0.8           & 35.8    \\
$1D$ $2^-$&2883& 0.0			& 0.0     &82.79    & 1.39	        & 132.0     \\
$1D$ $3^-$&2882& 34,7		& 0.3     &23.1     & 0.21          & 2.2 	\\
\hline
\end{tabular}
\end{center}
\end{table}

\section{THE $\bf{1^{--}}$ FAMILY}
All the states of this family of charmonium states can be clearly identify with a $J^{PC}=1^{--}$ value because they are produced via $e^+e^-$ annihilation.
\begin{table}
\caption{Predicted mass in MeV of the $J^P=1^-$ charmonium states diferent}
\label{t3}
\begin{center}
\begin{tabular}{|cccc|}
 \hline
 (nL) & States & QM & Exp. \\ 
 \hline
 (1S) & $J/\psi$ & $3096$ & $3096.916\pm0.011$  \\
 (2S) & $\psi(2S)$ & $3703$ & $3686.09\pm0.04$  \\
 (1D) & $\psi(3770)$ & $3796$ & $3772\pm1.1$  \\
 & $Y(4008)$ & $-$& $4008\pm40$ \\
 (3S) & $\psi(4040)$ & $4097$ & $4039\pm1$  \\
 (2D) & $\psi(4160)$ & $4153$ & $4153\pm3$  \\
 & $Y(4260)$ & $-$ & $4260\pm10$  \\
 (4S) & $\psi(4360)$ & $4389$ & $4361\pm9$  \\
 (3D) & $\psi(4415)$ & $4426$ & $4421\pm4$  \\
 (5S) & $\psi(4660)$ & $4641$ & $4664\pm1$  \\
 (4D) & $\psi(4660)$ & $4641$ & $4664\pm1$  \\[2ex]
\hline
\end{tabular}
\end{center}
\end{table}
Until the year 2005 five states appears as well established in the PDG \cite{PDG}, namely the $J/\psi$, $\psi(2S)$, $\psi(3770)$, $\psi(4040)$, $\psi(4160)$ and $\psi(4415)$. In that year a new state was reported by BaBar ~\cite{BB1}, the $Y(4260)$ decaying to $J/\psi\pi^+\pi^-$ soon confirmed by CLEO. Two years later while investigating whether the $Y(4260)$ decayed to $\psi(2S) \pi^+\pi^-$ BaBar discovered a new $1^{--}$ state: the $Y(4360)$ \cite{BB2}. Recently Belle has confirmed all these states and found two more. The first one Y(4660) ~\cite{BLL1} was not clearly visible in the BaBar data due to the limited statistic. The second appears when confirming the Y(4260)~\cite{BLL2}. The Belle experiment indicates that a fit using two interfering Breit-Wigner shapes describes the data better than using only the Y(4260). The new structure appears at $M=4008\pm 40^{+114}_{-28}$ MeV with a width $\Gamma=226\pm 44\pm 87$ MeV. Very recently a combined fit of the BaBar and Belle ~\cite{BBLL} data gives the best measurement of the $Y(4360)$ and $Y(4660)$ states. With respect to this last state a new peak with a significance of $8.8 \sigma$ has been observed \cite{Pak}  in the $\Lambda^+_c \Lambda^-_c$ invariant mass distribution of the $e^+e^-\rightarrow \Lambda^+_c\Lambda^-_c$ process with a mass and width of $M=4634\pm 8\pm 5$ MeV and $\Gamma=92\pm 40\pm 21$ MeV suggesting a new $1^{--}$ state in this energy region. 

In table ~\ref{t3} we compare the spectrum calculated with our model with the experimental data. As one can see the agreement with the experimental data is remarkable except for two states: the Y(4008) and the Y(4260) which does not fit in this scheme. In particular the Y(4360) is identified with the $4^3S_1$ charmonium excitation while we have two possible
assignement for the Y(4660) either $5^3S_1$ or $4^3D_1$. The other state may be the one identified in the $e^+e^-\rightarrow \Lambda^+_c\Lambda^-_c$ process.

We have also studied the leptonic decay width and the total strong decay width. They are show in table ~\ref{t4} and all of them are in agreement with the experimental data. It is generally assumed that to reproduce the leptonic widths it is necesary a sizeable mixture of $^3S_1$ and $^3D_1$ sates. In our model the mixing is not fitted but driven by the quark-quark interaction and is very small. Nevertheless we are able to reproduces the leptonic width because the flattening of our confinement potential.

\begin{table}
\caption{Strong and leptonic widths in keV of the $J^P=1^-$ charmonium states }
\label{t4}
\begin{center}
\begin{tabular}{|cccc|cc|}
 \hline
 Meson & Dominant Mode & $\Gamma_{the}$ (MeV) & $\Gamma_{exp}$ (MeV)& $\Gamma^{e^+e^-}_{the}$ &  $\Gamma^{e^+e^-}_{exp}$\\
 \hline
 $\psi(3770)$  &  $DD$              &  $22.2156$ & $26.3\pm1.9$& $0.22$ & $0.22\pm0.05$\\
 $\psi(4040)$  & $D^{\ast}D^{\ast}$ &  $92.9073$ & $80\pm10$& $1.11$ & $0.83\pm0.20$  \\
 $\psi(4160)$  & $D^{\ast}D^{\ast}$ &  $96.8348$ & $103\pm8$ & $0.30$ &  $0.48\pm0.22$ \\
 $\psi(4360)$  & $DD_{1}$           &  $89.8044$ & $103\pm11$ & $0.78$ &$-$\\
 $\psi(4415)$  & $DD_{1}$           & $133.1207$ & $119\pm16$& $0.33$ & $0.35\pm0.12$\\
 $\psi(4660)$  & $D^{\ast}D^{\ast}$ & $107.8655$ & $42\pm6$ & $0.31$ &$-$ \\
\hline
\end{tabular}
\end{center}
\end{table}

\section{THE $\bf 3940$ FAMILY}
Several states has been observed by different collaborations around the energy of $3940$ MeV (see ~\cite{God08} for a review). In table~\ref{t5} we show the results obtained with our model in this region. 
The Z(3930) was reported by Belle in $\gamma \gamma \rightarrow D \overline{D}$ with a mass and width $M=3929\pm 5\pm 2$ MeV and $\Gamma=29.9\pm 10\pm 2$ MeV \cite{ueh}. The two photon width is measured to be $\Gamma_{\gamma\gamma}B(Z(3930)\rightarrow D\overline{D})=0.18\pm 0.05\pm 0.03$ keV. Moreover the $D\overline{D}$ angular distribution is consistent with $J=2$.
The  $\chi_{c2}(2^{++})$ state is a good candidate for the Z(3930). We obtained a mass of 3968 MeV and the total width $\Gamma=49.1$ MeV is comparable with the experimental data. Finally the experimental two photon width compare nicely with our result
$\Gamma_{\gamma\gamma}B(Z(3930)\rightarrow D\overline{D})=0.15$ keV.

We can clearly identify a second state $\eta_{c2}(2^{-+})$ with mass $M=4166$ MeV and width $\Gamma=122.9$ MeV with the resonance recently reported by Belle at $M=4156^{+ 25}_{- 20}\pm 15$ MeV with a width $\Gamma=139^{+111}_{-61}\pm 21$ MeV ~\cite{bell} in the $e^+e^-\rightarrow J/\psi D^*\overline{D^*}$. The decay of $\eta_{c2}$ to $D\overline{D}$ is forbidden being the $X(4160)\rightarrow D^*\overline{D^*}$ the most favored decay channel as showed by the data.

The situation is more complicated for the rest of states. One can attempt to identify the state $\chi_{c1} (2P)\,1^{++}$ at M=3947 which decay into $DD^*$ with the $X(3940)$ which has been observed in the $X(3940)\rightarrow DD^*$ decay but there is no evidence of such state in the $D\overline{D}$ channel. However the existence of the X(3872) and the Y(3940) which cannot be described as simply $c\bar c$ states suggest that, as in the case of the $D_{sJ}$ mesons, the states with $J^P=0^+$ or $1^+$ may be mixed with tetraquark structures. A recent calculation ~\cite{hiy} with the same quark-quark interaction we use, shows that the $DD^*$ channel shows a very sharp resonance sligthly below the $D^0\overline{D}^{*0}$ threshold although the result seems to be rather sensitive to details of the quark-quark interaction. 
\begin{table}
\caption{Predicted mass in MeV for the diferent $c\overline s$ states in the 3940 MeV region }
\label{t5} 
\begin{center}
\begin{tabular}{|ccc|}
 \hline
 meson & $J_{PC}$  & mass(MeV) \\
 \hline
 $\eta_{c2}$   &  $(1S)2^{-+}$ & 3811\\
 $\chi_{c0}$  & $(2P)0^{++}$ &  3908  \\
 $\chi_{c1}$  & $(2P)1^{++}$ &3947 \\
 $h_c$  & $(2P)1^{+-}$&3955 \\
 $\chi_{c2}$  & $(2P)2^{++}$& 3968  \\
 $\eta_c$&$(3S)0^{-+}$ & 4054 \\
 $\eta_{c2}$&$(1D)2^{-+}$&4166 \\
\hline
\end{tabular}
\end{center}
\end{table}

\section{SUMMARY}
We have studied the properties of the new open and hidden charm mesons using a constituent quark model constraind by the light meson spectra. An important characteristic of the model is the use of an screened confinement potential which improve the description of the high excited meson states.
We confirm that the $D_{sJ}(2860)$ resonace has an important tetraquark component. Besides the well established $J/\psi$ and $\psi$ states, the $\psi(4360)$ and the $\psi(4660)$ can be considered as  $1^{--}$ charmonium states. However the  Y(4008) and the Y(4260) states can not be accommodated in the spectrum and probably its structure involves more complex configurations than a simple $c\bar c$ pair. In the 3940 Mev mass region we identify the Z(3930) as the $\chi_{c2} 2^3P_2 (c\bar c)$ state and the X(4160) as the $\eta_{c2} 1^3D_2(c\bar c)$ state. The properties of the rest of the states of this region are difficult to explain in a $q\bar q$ scheme.

\begin{acknowledgments}
This work has been partially funded by Ministerio de Ciencia y Tecnolog\'\i a
under Contract
No. FPA2007-65748, and by Junta de Castilla y Le\'on under Contract No. 
SA016A07.

\end{acknowledgments}

\end{document}